\documentclass[12pt,preprint]{aastex}
\def\lsim{\lower.5ex\hbox{$\; \buildrel < \over \sim \;$}}
\def\gsim{\lower.5ex\hbox{$\; \buildrel > \over \sim \;$}}
\lefthead{}
\righthead{  }

\def\lsim{\lower.5ex\hbox{$\; \buildrel < \over \sim \;$}}
\def\gsim{\lower.5ex\hbox{$\; \buildrel > \over \sim \;$}}

\begin{document}

\title{Generalized pseudo-Newtonian potential for studying 
accretion disk dynamics in off-equatorial planes
around rotating black holes: Description of a vector potential}

\author{Shubhrangshu Ghosh\altaffilmark{1} and Banibrata Mukhopadhyay\altaffilmark{1}}
\altaffiltext{1}{Astronomy and Astrophysics Program, Department of Physics,
Indian Institute of Science, Bangalore 560012, India; bm@physics.iisc.ernet.in} 

\begin{abstract}

We prescribe a pseudo-Newtonian vector potential for studying 
accretion disks around Kerr black holes. The potential is useful to study 
the inner properties of disk not confined to the equatorial plane
where general relativistic effect is indispensable. 
Therefore, we incorporate the essential properties of the metric at the inner radii through the pseudo-Newtonian potential derived from the general Kerr spacetime. The potential, reproducing most of the salient features of the general-relativity, is valid for entire regime of Kerr parameter. It reproduces the last stable circular orbit exactly 
as that in the Kerr geometry. It also reproduces last bound orbit and energy 
at last stable circular orbit with a maximum error $\sim 7\%$ and $\sim 15\%$ respectively
upto an orbital inclination $30^{\circ}$.    

\end{abstract}


\section{Introduction}

The standard model of the accretion disk
proposed by Shakura \& Sunyaev (1973) explains the dynamics of a thin disk by the Newtonian potential, not taking into account of the essential general-relativistic effects of the
black hole. The only notion it hires from general relativity (hereinafter GR) is that the inner most edge of the disk truncates at the last stable circular orbit of the Schwarzschild geometry. 
Complete general relativistic description of the thin Keplerian disk was developed by Novikov \& Thorne (1973). 
Owing to the complicated and cumbersome nature of underlying general relativistic equations, especially 
to study the full scale dynamics of the disk, it would be worthy enough to study in Newtonian
framework using a modified Newtonian 
potential, namely pseudo-Newtonian potential (hereinafter PNP), having all the essence of a black hole 
geometry. Such a potential was first 
proposed by Paczy\'nski \& Wiita (1980; hereinafter PW) for modeling a thick/thin accretion disk 
around a nonrotating black hole. The potential exactly reproduces marginally stable orbit $(r_s)$ 
and marginally bound orbit $(r_b)$ of that in full GR and the efficiency per unit mass at the last 
stable circular orbit $(E_s)$ agrees with at most $10 \%$ error. 
Nowak \& Wagoner (1991) and Mukhopadhyay \& Misra (2003) proposed other potentials for an accretion 
disk to describe epicyclic frequency which mimic most of the properties of the disk governed by the 
Schwarzschild and Kerr geometry respectively. 

PNPs to describe the fluid dynamics of 
accretion disk around a rotating black hole in equatorial plane were also
proposed by Chakrabarti \& Khanna (1992) and Artemova et al. (1996) which, however, were not valid for
entire range of Kerr parameter. Later,
a twodimensional potential was described by Mukhopadhyay (2002; hereinafter M02) which
is valid for the entire range and reproduces $r_s$ exactly, $r_b$ with a 
maximum error $\sim 5\%$ and $E_s$ with a maximum possible error $\sim 10\%$ 
compared to that in the Kerr geometry. 
The more interesting fact, however, lies in the methodology adopted by 
M02 in deriving the potential which can be used to formulate 
the PNP for any metric according to the physics concern. Based on the 
procedure, several authors have  developed PNPs at the equatorial plane for other metric, e.g 
Hartle-Thorne metric by Ghosh (2004), Kerr-Newman metric by Ivanov \& Prodanov (2005). We plan to 
follow the procedure outlined by M02 to derive a PNP for general Kerr geometry (particle motion not 
confined to the equatorial plane) which can be applied to study the generalized axisymmetric
accretion disk dynamics. Naturally, the aim is to furnish a pseudo-Newtonian vector potential
rather than a scalar potential with radial and vertical component in cylindrical coordinate
system suitable for studying a thick accretion disk.

While some authors focus on to propose an ad-hoc and a mathematically simpler form of PNP, 
we derive it from the spacetime metric directly. Therefore, we do not need to worry about its
final form strictly determined by our procedure for any spacetime metric 
unlike the other PNPs. Problem with an ad-hoc PNP is that there is no easy one to one correspondence 
between spacetime metric and PNP.

Most of the cosmic objects in our nature are rotating (see e.g. Iwasawa et al. 1996; Wilms et al.
2001; Miniutti et al. 2004; Miller et al. 2004; Genzel et al. 2003). Recently, it has also been
argued (McClintock et al. 2006) that the source GRS~1915+105 is a rapidly rotating black hole
with spin parameter $a>0.98$. Thus, the predictions of 
the inner disk properties will be incorrect if the pseudo-Newtonian modeling fails to include 
the spin parameter of the black hole. 
The potential proposed by M02 has already been used to study the hydrodynamics of the disk around Kerr black holes, which shows a dramatic change of the valid parameter space region; any shock location 
and other fluid properties change for different rotation parameters of the black hole 
(Mukhopadhyay 2003). Recently Chan et al. (2005, 2006) have used this potential for time dependent 
studies of accretion flow for twodimensional viscous hydrodynamic disk and for threedimensional 
self-gravitating magnetohydrodynamic disk around rotating black holes using spectral method. 

Our aim in the present paper is to develop a potential for the general Kerr geometry which can be 
used to mimic the general relativistic effects of the spacetime, enabling us to suitably use it to 
study the accretion dynamics, especially of a thick 
disk with the inclusion of vertical outflow. Thus we goal for a generalized potential which 
is a function of $r$ and $\theta$ both and having two components, in radial and vertical direction, 
in cylindrical coordinate system, hence a vector potential.  
This is useful for both quasi-numerical analyzes and 
time dependent simulations, avoiding the use of complicated  relativistic equations, yet capturing 
the essential physics of spacetime outside of a Kerr black hole. In early, some attempt was made to
describe PNP in generalized Kerr geometry by, e.g.,
Dadhich (1985), Semer\'ak \& Karas (1999), Chakrabarti \& Mondal (2006). The PNPs described
in first two works, while aim to furnish frame-dragging effects in the Newtonian framework, do
not reproduce marginally stable and bound circular orbits and specific energy close to the black hole
and hence may not be suitable for the hydrodynamical analysis of 
a Keplerian accretion disk, particularly, when aim is to study inner disk
properties. The third one is the modification of Chakrabarti \& Khanna (1992) potential to mimic 
approximately the general relativistic effects in accretion disks 
for $-1 \leq a \leq 0.8$. The potential is a prescription 
with which the authors maintain the ad-hoc-ness of PW potential and may be used to study
accretion disk having particle orbits of small inclination angle at a certain parameter window.
Our goal here, however, is to extend the procedure initiated by 
 M02 which can be used to describe a PNP for the general Kerr spacetime; 
particle motion not confined to the equatorial plane. The potential given by M02 is more 
useful to study the dynamics of a vertically averaged thin disk at equatorial plane. Henceforth, we wish 
to derive a generalized potential applicable for off-equatorial planes 
following the procedure of M02 which reproduces exactly or in good
agreement all the (inner) accretion disk properties (unlike the previous works) for a general case in 
Kerr geometry. This will be useful to study the thick accretion disk dynamics 
with the inclusion of vertical outflow and wind. The potential should reproduce those features of a 
rotating black hole geometry which have been reproduced by that of M02 at equatorial plane and that of 
PW for a nonrotating black hole. We formulate our PNP from the Kerr metric. As the metric is involved 
directly to our calculation, many of the features of Kerr geometry are inherent in our potential. 
The axisymmetric PNP we plan to establish here for a general Kerr metric is more useful for 
quasi-analytical studies and time dependent simulations of accretion and wind and related
physics as that studied by Chan et al. (2005, 2006), Lipunov \& Gorbovskoy (2007), Shafee et al. (2007) 
using M02, without loss of any generality. 

We arrange the paper in the following manner. In the next section, we present the basic equations and derive the PNP. In \S 3, we compare the various features of dynamics of accreting particles obtained by our potential with that in Kerr geometry. Finally \S 4 presents a discussion with summary. 

\section{Formulation of the generalized pseudo-Newtonian potential}

Here we furnish the derivation of our said potential. It is condensed 
in three subsections, where we have elucidated the description of the potential starting 
from the Kerr metric. 

\subsection{Integrals of motion of a test particle in the Kerr geometry}

The Kerr spacetime in the Boyer-Lindquist coordinate system in geometrized units $G = c = 1$ is given by  
\begin{eqnarray}
\nonumber
ds^2=-\left(1-\frac{2Mr}{\Sigma}\right)dt^2-\frac{4aMr \sin^2 \theta}{\Sigma}{dt} {d\phi}
+\frac{\Sigma}{\Delta}dr^2\\+\Sigma d\theta^2+\left(r^2+a^2+
\frac{2Mra^2 \sin^2 \theta}{\Sigma}\right) \sin^2 \theta \, d\phi^2,
\label{sag2}
\end{eqnarray}
where $\Delta=r^2+a^2-2Mr$, $\Sigma=r^2+a^2 \cos^2 \theta$ and $a=\frac{J}{M}$ 
which is specific angular momentum of the black hole named Kerr parameter. The Lagrangian density of the particle of mass $m$ in the Kerr spacetime
is given by
\begin{eqnarray}
2 {\cal L}=-\left(1-\frac{2Mr}{\Sigma}\right){\dot t}^2-\frac{4aMr \sin^2 \theta}{\Sigma}{\dot t} {\dot \phi}
+\frac{\Sigma}{\Delta}{\dot r}^2 \nonumber \\+\Sigma {\dot \theta}^2+\left(r^2+a^2+
\frac{2Mra^2 \sin^2 \theta}{\Sigma}\right)\sin^2\theta {\dot \phi}^2,
\label{sag4}
\end{eqnarray}
where over-dots denote the derivative with respect to proper-time $\tau$ along the curve. 
From the symmetries, we obtain two constants of 
motion corresponding to two ignorable coordinates $t$ and $\phi$ given by
\begin{eqnarray}
p_t=\frac{\partial \cal L}{\partial \dot t} = -\left(1-\frac{2Mr}{\Sigma}\right) \dot t-
\frac{2aMr \sin^2 \theta}{\Sigma} \dot \phi ={\rm constant}=-E 
\label{12}
\end{eqnarray}
and 
\begin{eqnarray}
p_\phi=\frac{\partial \cal L}{\partial \dot \phi}=-\frac{2aMr \sin^2 \theta}{\Sigma} \dot t +\left(r^2+a^2+
\frac{2Mra^2 \sin^2 \theta}{\Sigma}\right)\sin^2\theta \dot \phi={\rm constant}=\lambda_z, 
\label{13}
\end{eqnarray} 
where $\lambda_z$ is the component of angular momentum of the orbiting particle
along spin axis of the black hole (i.e. parallel to the symmetry axis).
The Hamiltonian of the system can also be written as 
\begin{eqnarray}
{\cal H} =\frac{1}{2} g^{\mu\nu} p_{\mu} p_{\nu}.
\label{15}
\end{eqnarray}
The Hamiltonian is itself a constant of motion and by usual setting $g_{\mu\nu} p^{\mu} p^{\nu}=-m^2$
for a particle with nonzero rest mass, we obtain the third constant of motion (integral of motion) as
\begin{eqnarray}
{\cal H} = \frac{1}{2} m^2={\cal L}=\rm{constant}.
\label{16}
\end{eqnarray}
The Lagrangian and the Hamiltonian of the geodesic with the above condition are exactly same. The three 
integrals of motion obtained above are sufficient to determine the geodesic equations under 
restrictive condition like the motion of a test particle confined to an 
equatorial plane $(\dot \theta=0, \theta=\pi/2)$ (e.g. Shapiro \& Teukolsky 1983). To describe the motion of a particle for a generic case around the Kerr black hole, with dependency on $\theta$, a fourth integral of motion is needed which does not come from the symmetries of the metric. It was 
discovered by Carter (1968) that such an integral of motion can be obtained explicitly by separability
of the Hamilton-Jacobi equation (hereinafter HJE) (also see Chandrasekhar 1983) given by  
\begin{eqnarray}
Q  =  p^{2}_\theta + \cos^2\theta\,[a^2(m^2 - E^2) + \frac{\lambda^{2}_z}{\sin^{2} \theta}].
\label{23}
\end{eqnarray}
$Q$ is called the Carter constant (Carter 1968). The above equation together with eqns. (\ref{12}) 
and (\ref{13}) provides a complete set of integrals of motion for the geodesic of a test particle in 
the Kerr geometry where $Q$, $E$ and $\lambda_z$ are three constants motion. 
It can be seen 
from the above equation that for a particle always confined to the equatorial plane (i.e., $\dot \theta = 0$
and $\theta = \pi/2$), the Carter constant $Q$ is zero. 
For $a=0$, $Q + \lambda^{2}_z=\lambda^2$ is square of the total angular 
momentum of a particle's orbital motion. 
It is to be noted that the information about $\theta$ and $\dot \theta$ for an orbital trajectory 
is consolidated into a single parameter $Q$, which is a conserved quantity. Now
using eqns. (\ref{sag4}), (\ref{12}), (\ref{13}), (\ref{16}) and (\ref{23}),  
 we obtain the most general equations of motion governing the orbital trajectory of a test particle in the Kerr geometry
(Bardeen 1972; Wilkins 1972) given by
\begin{eqnarray}
\Sigma \frac{dr}{d \tau}=\pm \sqrt{R}, 
\label{24}
\end{eqnarray}
\begin{eqnarray}
\Sigma \frac{d \theta}{d \tau}=\pm \sqrt{\Theta},
\label{25}
\end{eqnarray}
\begin{eqnarray}
\Sigma \frac{d \phi}{d \tau}=- a E + \frac{\lambda_z}{\sin^2 \theta}
    + \frac{a}{\Delta}[E(r^2+a^2) - \lambda_z a], 
\label{26}
\end{eqnarray}
\begin{eqnarray}
\Sigma \frac{dt}{d \tau}=- a^2 E \sin^2 \theta + a \lambda_z +
\frac{r^2 + a^2}{\Delta} [E(r^2 + a^2)-\lambda_z a],
\label{27}
\end{eqnarray}
where
\begin{eqnarray}
R=[ E(r^2 + a^2) - \lambda_z a]^2 - \Delta \, [m^2 r^2 + (\lambda_z - a E)^2 + Q],
\label{28a}
\end{eqnarray}
\begin{eqnarray}
\Theta=Q - \cos^2\theta\,[a^2(m^2 - E^2) + \frac{\lambda^{2}_z}{\sin^{2} \theta}].
\label{28}
\end{eqnarray}

\subsection{Effective potentials and prescription of orbital inclination}

\begin{figure}
\epsscale{0.8}
\hskip-5cm
\plotone{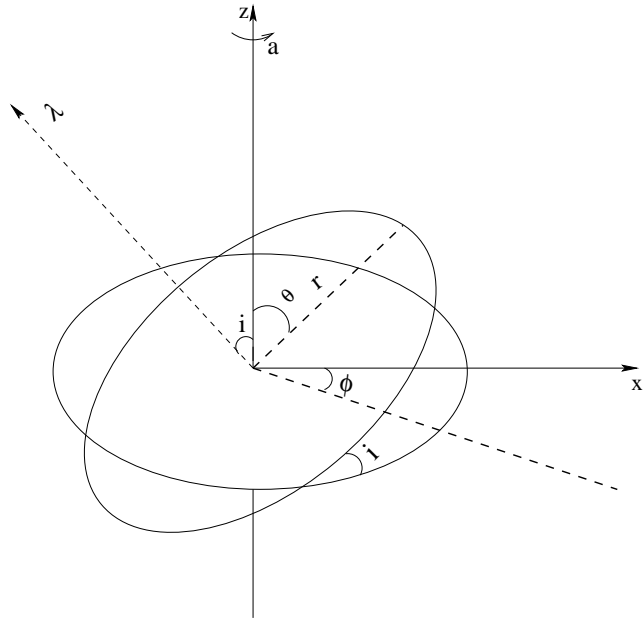}
\caption{ \label{fig1}
Schematic diagram showing how an orbit of the
accreting particle inclines to the equatorial
plane. The orbit, inclined at an angle $i$,
represents a generalized accretion.
$\lambda$ represents the total angular momentum of the
orbiting particle, $a$ is the spin parameter of the
black hole and $z$ axis is the symmetry axis of the system.
}
\end{figure}

$R$ and $\Theta$ are in general called the effective potentials describing the particle motion 
in a generalized Kerr spacetime. 
For a particle moving perpetually at the equatorial plane, 
the only effective potential that governs the 
particle motion is $R$ with $Q$ equals to zero. In order to understand the generic orbit, both potentials should be accounted for. To have an insight of the nature of 
orbits in threedimensional Kerr geometry, we recall that for bound geodesics $(Q \geq 0)$ orbits 
either remain perpetually at the equatorial plane $(Q=0)$ or cross it repeatedly $(Q > 0)$. 
More precisely, orbits with $Q > 0$ (more general case), also called the off-equatorial orbits, are inclined to the equatorial orbital plane. Figure \ref{fig1} shows that the orbital plane of an off-equatorial orbit is inclined at an angle $i$ to the equatorial plane, which also means that the 
orbital plane makes an inclination $i$ with the spin axis of the hole (symmetry axis). For a particular 
orbit, the inclination angle $i$ is a constant of motion.
However, in general, the plane of an off-equatorial orbit precesses with an angular frequency $2 a/r^3$ called Lense-Thirring precession (Lense \& Thirring 1918). Owing to this fact, the plane of the orbit neither remains fixed, nor crosses the equatorial plane
with constant inclination $i$, but rather at a different angle. Looking meticulously, we find that 
the information 
of entire $\theta$ motion (as for a particular off-equatorial orbit, $\theta$ has 
all arbitrary values and it changes constantly) is glued up into 
one single inclination parameter $i$. Thus the motion of the bound orbit in the 
Kerr geometry is now governed by four parameters $r$, $i$, $\phi$ and $t$.

As our intention is to derive a generalized PNP for particle orbits not confined to the
equatorial plane, which itself is an approximate quantity, we restrict ourselves to an ideal case of 
circular orbits. In general, due to plane precession, the orbit never remains circular, but have eccentricity. The study 
of generic orbits (inclined and eccentric) for our particular case is deferred for future.  The circular 
orbits satisfy $R=dR/dr=0$. 
Hence to analyze the orbital motion for circular case, the radial effective potential $R$ is necessary and sufficient. 

Our next objective is to describe the inclination parameter $i$ with known quantities. As stated earlier, the constant inclination 
angle $i$ (without considering plane precession) is the angle between the plane of the orbit and the 
spin axis of the hole. 
As $\lambda^2=Q+\lambda_z^2$ for $a=0$ stated in the preceding subsection, 
we identify in the Cartesian coordinate system
$Q \sim \lambda^{2}_x + \lambda^{2}_y$ approximately.
Hence $Q$ can be treated as the component of square of the total angular momentum in
perpendicular to the symmetry axis or to the spin axis ($\lambda_\perp$)
of the black hole or precisely the
component projected into the equatorial plane.
However, even
for a maximally spinning hole the error is found to be less than a few percent (Glampedakis et al. 2002). With this conjecture, the inclination parameter $i$ can be defined as (Ryan 1995; Shapiro 1996; Hughes 2000)
\begin{eqnarray}
\cos i = \frac{\lambda_z}{\sqrt{Q+\lambda^{2}_z}}.
\label{sa1}
\end{eqnarray}
Taking all these into account, our radial effective potential $R$ in eqn. (\ref{28a}) 
with further simplification eventually reduces to 
\begin{eqnarray}
R=r(r^3+ra^2+2Ma^2)E^2 - r(r-2M)\lambda^{2}_z - 4aMr\lambda_{z} E 
- \Delta\, m^2r^2 - \Delta\, \lambda^{2}_z \tan^{2} i.
\label{shuR}
\end{eqnarray}
Thus the effective potential is a function of $r$ and $i$ (which is a conserved quantity for a particular orbit). If we consider the effect of Lense-Thirring precession on the orbital plane, then 
we need to amend a minor correction with the above definition of inclination angle.             
Now if you specify for circular orbits, then the effective potential further reduces as
\begin{eqnarray}
\nonumber
V_{\rm{eff}}= \frac{2aM\lambda_z + \biggl[\lambda_z^2\{r^2 + \frac{\tan^2 i}{r} (r^3 + ra^2 + 2Ma^2)\}
\Delta + m^2 r(r^3 + ra^2 + 2Ma^2)\Delta \,\sec^2 i \biggr]^{1/2}}{(r^3 + ra^2 + 2Ma^2)\sec i}\\
\label{shu6}
\end{eqnarray}
which is essentially the solution for $E$ of $R=0$.

\subsection{Description of pseudo-Newtonian force}

The conditions for a circular orbit are 
\begin{eqnarray}
R=0, \hskip1.cm\frac{dR}{dr}=0.
\label{sa2}
\end{eqnarray}
Solving for $E$ and $\lambda_z$ from eqn. (\ref{sa2}) we obtain
\begin{eqnarray}
\tilde{E}=\frac{E}{m}=F_1 (r, a, i) \, \, \, {\rm{and}}\, \, \, {\tilde{\lambda}}_z=\frac{\lambda_z}{m}=F_2 (r, a, i).
\label{sa3}
\end{eqnarray}
$\tilde{E}$ and ${\tilde{\lambda}}_z$ are specific energy and 
specific angular momentum parallel to the symmetry axis 
respectively, and $r$ is the radius of the circular 
inclined orbit. We do not represent the functions $F_1$ and $F_2$ explicitly as they are cumbersome. 
 It can be easily verified that $\lambda_z \sec i=\lambda$ is 
the total angular momentum of the orbital plane with 
inclination angle $i$ and its direction is perpendicular to 
the plane of the orbit. The relativistic specific angular 
momentum of the matter is defined by the 
relation $\ell=- u_{\phi}/u_t= \lambda_z/E$ (Kozlowski et al. 1978). Following 
M02 and as standard practice, we can define the Keplerian 
angular momentum distribution $\lambda_K=\ell \sec i$. Therefore, corresponding 
centrifugal force in the Kerr geometry can be written as 
\begin{eqnarray}
\frac{{\lambda_K}^2}{r^3} & = & \frac{2M{\cal A}^2 \sec^2 {i}}{\biggl[a\sqrt{2} \sqrt{M} r^{3/2}\{\Delta+2r(r-M)\}+
r\Delta\, \Bigl\{[{\cal A}+r^4-a^2(\Delta+r^2-3Mr)]\cos {2i} \sec^2 {i}\Bigr\}^{1/2}\biggr]^2} \nonumber  \\
& = & F_G =F_{Kr}
\label{sa5}
\end{eqnarray} 
where ${\cal A}=a^4 + r^4 + 2a^2r(r-2M)$. Thus from above, $F_{Kr}$ can be 
identified as the generalized gravitational force  of the Kerr black hole at the Keplerian 
orbit. The subscript $K$ represents a Keplerian orbit and 
subscript $r$ represents force in the radial direction. 
The above expression reduces to that of M02 at an inclination angle $i=0$, i.e. particle 
motion confined to the equatorial plane, and that of PW for $a=i=0$. Thus 
we can say that above eqn. (\ref{sa5}) is the most general form 
of force corresponding to the PNP for a generalized 
accretion disk around a rotating black hole. One may be confused with the mathematical form of $F_{Kr}$
in eqn. (\ref{sa5}) which appears complicated compared to that of PW and thus the PNP
approach with $F_{Kr}$ appears to be 
lacking the main motivation and advantage in comparison with full framework of GR.
However, one should also remember that mathematical complexity of accretion disk dynamics in
full Kerr geometry is beyond comparison with that in much simpler Schwarzschild geometry. Therefore,
the apparent complex form of $F_{Kr}$ above is nothing compared to the complications exist
in the set of actual equations in the Kerr geometry. Therefore, $F_{Kr}$ must be acceptable
if it exhibits the essential features of the Kerr geometry what we examine in \S3. 

\begin{figure}
\epsscale{1.2}
\hskip-1.0cm
\includegraphics[width=1.2\columnwidth,angle=270]{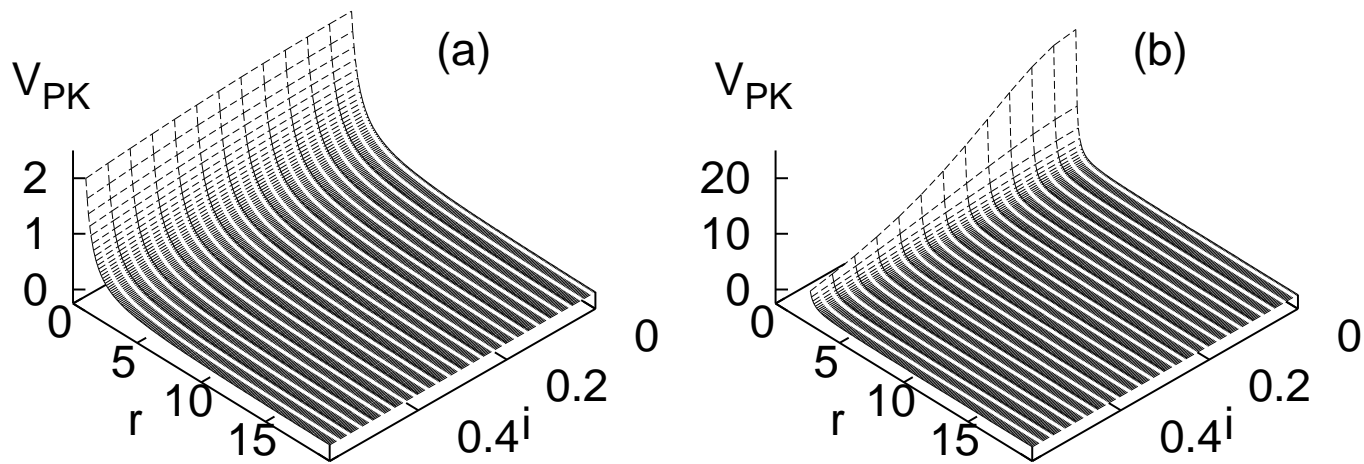}
\caption{ \label{fig2}
Variation of generalized pseudo-Newtonian potential $V_{PK}$ 
as a function of $r$ (in units of $GM/c^2$) and $i$ (in radian) 
for (a) $a=1$, (b) $a=-1$.
}
\end{figure}

\begin{figure}
\epsscale{0.8}
\plotone{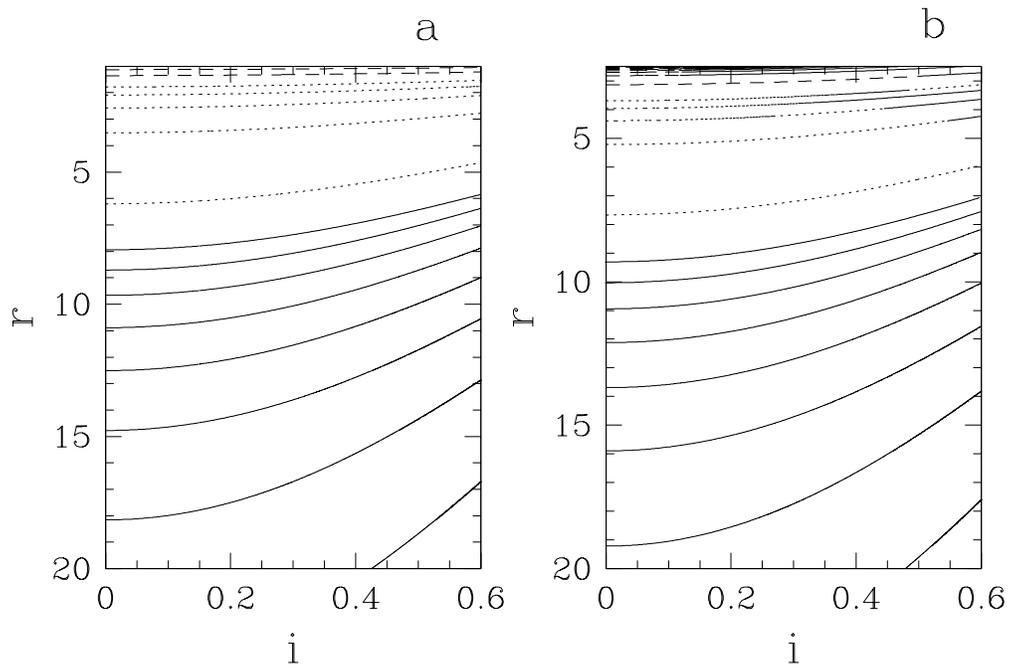}
\caption{ \label{fig3}
Contours of constant $V_{PK}$ for
(a) $a=1$, when solid lines correspond to $V_{PK}=0.045,0.06,0.075,......,0.15$,
dotted lines to $0.2,0.4,0.6,......,1$, and
dashed lines to $1.5,2$,
(b) $a=-1$, when solid lines correspond to $V_{PK}=0.045,0.06,0.075,......,0.15$,
dotted lines to $0.2,0.4,0.6,......,1$, and
dashed lines to $2,4,6,......,10$.
}
\end{figure}

The general form of corresponding PNP is 
\begin{eqnarray}
V_{PK}=\int F_{Kr} dr. 
\label{3dpotk}
\end{eqnarray}
Therefore, $V_{PK}$ is a generalized axisymmetric potential mimicking the geometry of 
the general Kerr spacetime. The above potential can be 
attributed to the threedimensional free-fall acceleration 
of a test particle around a Kerr black hole. In Fig. \ref{fig2} we show the
evolution of the pseudo-Newtonian potential as a function of $r$ and $i$ for
extremally rotating black holes. As the unstable region including unbound orbits with small
inclination is broader for a counter rotating black hole (described in detail in \S 3 with
Tables 1 and 2), $V_{PK}$ increases and tends
to diverge at a larger radius than that of the corotating one. For the corotating case,
$V_{PK}$ tends to diverge at a very inner region in the vicinity of the black hole only. 
Figure \ref{fig3} describes the corresponding contours of constant $V_{PK}$.
It is found that the entire family of contour is shifted towards smaller $r$ for 
the corotating case compared to the counter rotating one. Above qualitative features are
same for other Kerr parameters.

Now we write the Keplerian force components in the radial and vertical direction in cylindrical 
coordinate system $(\rho,\phi,z)$ from eqn. (\ref{sa5}) as 
{\small
\begin{eqnarray}
\hskip-1.5cm
F_{K\rho} = \frac{2M{\cal A}^2 {\cal B}^{-1/2}}{\left[a\sqrt{2M} \rho^{3/2} {\cal B}^{1/4}\{\Delta+
2\rho {\cal B}^{1/2}(\rho {\cal B}^{1/2}-M)\}+
\Delta \sqrt{\rho^2-z^2} \left\{{\cal A}+\rho^4 {\cal B}^2 -a^2(\Delta+
\rho^2 {\cal B} -3M\rho {\cal B}^{1/2})\right\}^{1/2}\right]^2}     \nonumber \\
\label{sa8}
\end{eqnarray}
}
and
{\small
\begin{eqnarray}
\hskip-1.5cm
\nonumber 
F_{Kz}=\frac{2M{\cal A}^2 {\cal B}^{-1/2}z}{\left[a\sqrt{2M} \rho^{2} {\cal B}^{1/4}\{\Delta+
2\rho {\cal B}^{1/2}(\rho {\cal B}^{1/2}-M)\}+\Delta \rho^{1/2}\sqrt{\rho^2-z^2} \left\{{\cal A}+
\rho^4 {\cal B}^2 -a^2(\Delta+\rho^2 {\cal B} -3M\rho {\cal B}^{1/2})\right\}^{1/2}\right]^2}\\
\label{sa9} 
\end{eqnarray}
}
respectively, where ${\cal B}= 1+z^{2}/\rho^2$, $\Delta=\rho^2{\cal B}-2M\rho{\cal B}^{1/2}+a^2$,
${\cal A}=a^4+\rho^4{\cal B}^2+2a^2\rho{\cal B}^{1/2}(\rho{\cal B}^{1/2}-2M)$. 
$F_{K\rho}$ and $F_{Kz}$ are the components of force corresponding to the pseudo-Newtonian
vector potential components in cylindrical geometry.

\begin{figure}
\epsscale{1.2}
\hskip-1.0cm
\includegraphics[width=1.2\columnwidth,angle=270]{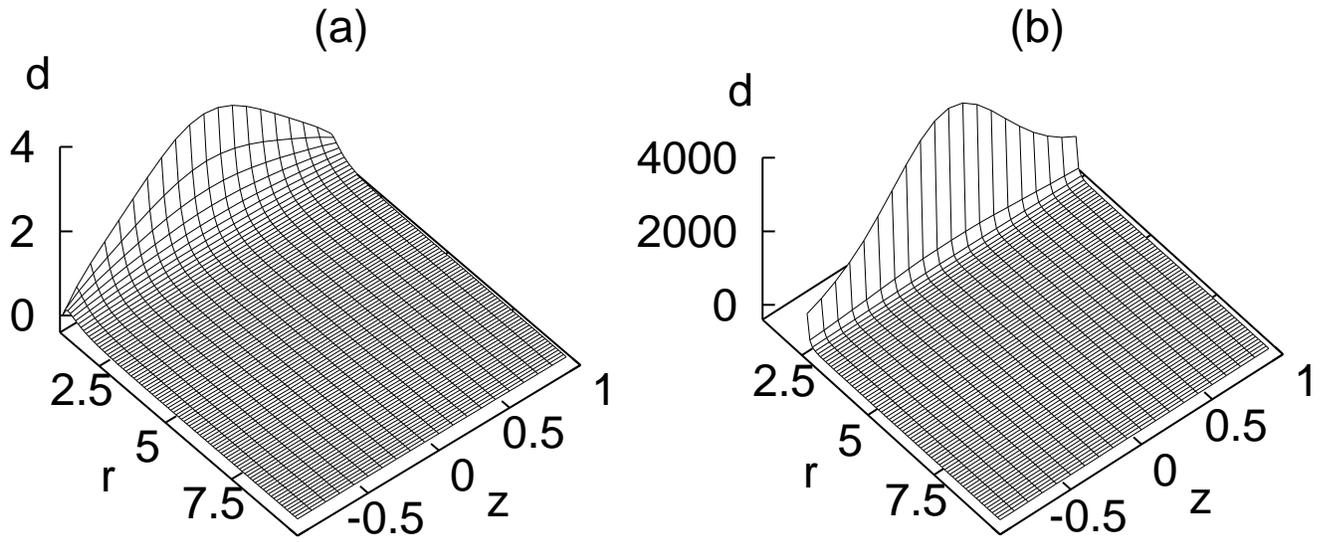}
\caption{ \label{fig4}
Variation of density distribution generating $V_{PK}$ 
as a function of $r$ and $z$ (in units of $GM/c^2$) for (a) $a=1$, (b) $a=-1$.
}
\end{figure}

Once we have the pseudo-Newtonian potential (and force), we 
examine the corresponding density distribution $d$ from the Poisson's equation
\begin{eqnarray}
-4\pi d =\frac{1}{\rho}\frac{\partial}{\partial \rho}\left(\rho F_{K\rho}\right)
+\frac{\partial}{\partial z} F_{Kz}
\label{hypden}
\end{eqnarray}
shown in Fig. \ref{fig4} for extremally rotating black holes. It is found that the
hypothetical density distribution $d$ is Gaussian along $z$ axis for both corotating
and counterrotating black holes. Like $V_{PK}$, $d$ increases much faster at an outer radius 
for the counterrotating black hole than the corotating one. $d$ for the corotating case 
increases significantly only at very close to the black hole event horizon.



\section{Comparison of the essential features for the Kerr geometry and pseudo-Kerr}

To check the validity of the potential derived in the preceding 
section, we follow the procedure considered in M02 and Mukhopadhyay \& Misra (2003). 
We make a comparison to show how far the 
potential $V_{PK}$ reproduces the values of $r_s$, $r_b$ and 
$E_s$ as that in Kerr geometry. We do not repeat the corresponding 
equations to calculate the parameters in Kerr geometry as 
they are established well in literature (see e.g. Shapiro \& Teukolsky). 
However, for completeness, we show them for the pseudo-Kerr case. 
Stability of an orbit requires (PW)
\begin{eqnarray}
\frac{d}{dr} (\lambda_K) \geq 0. 
\label{shu2}
\end{eqnarray}
Therefore, we obtain marginally stable orbit $(r_s)$ by 
solving eqn. (\ref{shu2}) for $r$ with ``equals to" sign. 

\begin{figure}
\epsscale{0.8}
\plotone{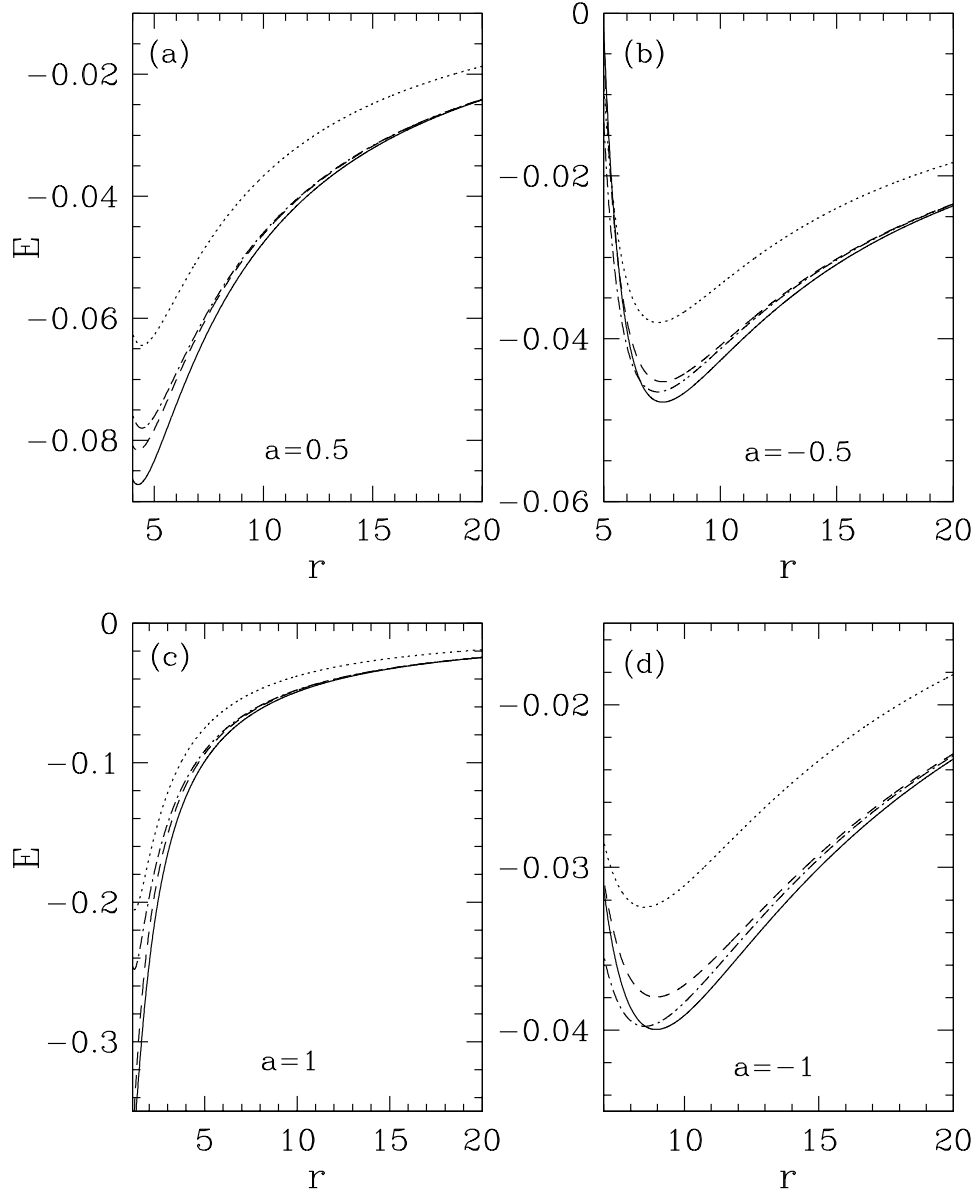}
\caption{ \label{fig5}
Variation of specific energy for circular orbits as a function of radial coordinate at 
various Kerr parameters.
The solid and dotted curves are for our pseudo-Newtonian potential with $i=10^\circ$ and $30^\circ$
respectively and the dashed and dot-dashed curves are for Kerr geometry with $i=10^\circ$ and $30^\circ$
respectively at each panel.
}
\end{figure}

The energy of a particle in the Kerr geometry according to $V_{PK}$ is given by
\begin{eqnarray}
E=\frac{r}{2} \frac{dV_{PK}}{dr} + V_{PK}.
\label{shu4}
\end{eqnarray}
Therefore marginally bound orbit ($r_b$) can be accounted for by solving the equation
\begin{eqnarray}
E\vert_{r_b}=0
\label{shu5}
\end{eqnarray}
for $r_b$.
In addition, once we know $r_s$, we can compute $E\vert_{r_s}=E_S$, 
the energy at last stable circular orbit.
We enlist the comparison of $r_s$, $r_b$ and $E_s$ between the Kerr 
and pseudo-Kerr geometry with percentage error for a few inclination 
parameters $i$ in the Tables 1,2,3. We present the values of above 
parameters upto the orbital inclination $i=30^\circ$, as we believe that realistically 
the disk thickness does not exceed beyond the 
height corresponding to inclination parameter $i \sim 30^\circ$. 
It is exciting to note that $r_s$ computed from pseudo-Kerr 
geometry exactly matches to that in Kerr geometry for all 
values of $a$ and $i$. To best of our knowledge, none of the existing pseudo-Newtonian potentials
proposed for threedimensional Kerr geometry reproduces $r_s$ exactly.
Table 1 shows that values of $r_b$ obtained from $V_{PK}$ match with that 
from exact Kerr geometry in very good agreement; with a 
maximum error $\sim 7\%$ at high inclination angle $30^{\circ}$. 
With the increase of $a$ upto certain value, error increases to a maximum
(obtained in the range $0.998\ge a \ge 0.9$) and then decreases to $0$ at $a=1$. 
It is to be noted that values of $r_s$ and $r_b$ are independent of $i$ at $a=1$ which
coincide with the event horizon as the case for $i=0$. 
The energy of an accreting particle at marginally stable 
orbit is also being reproduced by our PNP in very 
good agreement with that in exact Kerr 
geometry for $i \lesssim 20^{\circ}$, as understood from Table 2. 
Although the error increases with the increase of $i$ beyond $20^\circ$, except for $a=1$ which is
a special case, it is still within $20\%$  for any realistic disk. 
For counter-rotating black holes the 
errors are less than that for co-rotating ones. However, it is 
to be noted that for very high inclination angle this 
trend reverses in $E_s$ as shown in Fig. \ref{fig5} and Table 2. In both the cases of Kerr and 
pseudo-Kerr geometry, the orbits recede with the 
increase of inclination parameter $i$ for prograde black holes, 
and for retrograde black holes this dynamics is opposite. We compare the 
variation of specific energy for circular Keplerian orbit with 
the radial distance in Fig. \ref{fig5} computed for exact Kerr geometry 
and our PNP for both prograde and retrograde holes at 
orbital inclination $10^{\circ}$ and $30^{\circ}$. 

The pseudo-Newtonian potential described by Semer\'ak \& Karas (1999),
while tries to capture frame-dragging effects, can not reproduce other general relativistic 
properties e.g. radii of marginally stable and bound orbit, energy at marginally stable 
orbit. These are very important for the purpose of accretion disk producing jet,
expected to be launched from inner edge of the disk. On the other hand, our 
potential is derived directly from the spacetime metric, more precisely from the
effective potential of the threedimensional Kerr geometry given by eqns. (\ref{shuR}) and (\ref{shu6}).  
Either of the equations inherently carries information of 
frame-dragging effects with the term proportional to $a\lambda_z$. Therefore, the 
frame-dragging effects are expected to be guaranteed by our potential.
While the ad-hoc potential given by Chakrabarti \& Mondal (2006) tries to mimic frame-dragging
as well as inner disk properties both, it is not valid for entire parameter regime.
It also does not reproduce the radii of marginally stable orbit exactly as ours does. 
While their potential is a function of $r$ and $\theta$ both, they discuss most of the results at
equatorial plane ($\theta=\pi/2$) and thus it is not clear how effective their potential is in 
a real system. Therefore, our potential is expected to be most selfcontained
and useful to study generalized accretion disk including jet, particularly for the 
hydrodynamical/magnetohydrodynamical analysis.

Thus, for a realistic accretion disk, our PNP, 
although approximate in nature, should reproduce the essential features 
of the spacetime in very good agreement. The potential is 
effectively useful to study the detailed hydrodynamics of 
the realistic disk with the inclusion of vertical outflow using
Newtonian MHD equations, especially in the inner region, where 
general relativistic effects are very important. Thus 
the potential satisfies all the criteria to be claimed as 
good pseudo-Newtonian vector potential for a general Kerr metric.



\section{Discussion}

In this paper, we have prescribed a most generalized 
gravitational potential, pseudo-Newtonian vector potential, for the modeling of 
thick accretion disks around rotating 
black holes.  As this potential is derived from the 
spacetime metric directly, it naturally carries 
several essential general relativistic properties. 
Our potential is valid for co-rotating as well as counter-rotating black holes with 
all realistic values of orbital inclination. The inherent 
nature of reproducing exact values of $r_s$ for the entire 
regime of Kerr parameter and for all inclination angle 
by our PNP makes it an efficient choice. The prescribed PNP 
also reproduces $r_b$ in exact Kerr geometry with a 
maximum error $\sim 7\%$ for a high inclination $\sim 30^{\circ}$. 
However, it can be noted that the potential may not be a good approximation for the angular and 
epicyclic frequencies for less than $20$ Schwarzschild radii.

The full scale disk dynamics is very complicated and there is no virtue 
to assume that the accretion disk around black holes is 
fully Keplerian in nature. Moreover, the vertical outflow and 
wind in the form of relativistic jet, which are believed to be 
present almost in all quasars and microquasars, are predicted to 
be emanating from a very inner region of a puffed up disk or from a 
coronal transition region (Liu et al. 1999; Meyer et al. 2000). In addition, 
various inner fluid dynamical properties and also the powering of jet 
depend on spin parameter of the black hole. Thus evidently the realistic 
accretion physics (especially the accretion dynamics) around black 
holes can be studied with our prescribed axisymmetric generalized PNP, capturing the 
essential physics of GR. 

The general relativistic study of full scale dynamics of a realistic accretion disk 
for a highly turbulent viscous medium with the inclusion of jet 
and magnetic field is beyond the scope with the present physical 
and mathematical tools available to us. Most of the numerical studies 
of general relativistic flows have been performed with simplifying 
assumptions and remained confine to certain regions. 
Such simulations do not explicitly disclose the underlying 
microphysics of the structure and dynamics of disks and jets. In order to 
avoid complexities in full GR frameworks and with a notion to understand the underlying 
physics of the symbiotic connection of disk and jet, our 
PNP is highly recommended. Even if the results using our PNP deviate 
quantitatively by a few percent from that of the actual Kerr geometry, the 
accurate physical interpretation of the activities around black holes 
is possible. Our generalized PNP can be used to perform
quasi-analytical studies and time-dependent simulations of accretion and 
wind and related physics like that done by Chan et al. (2005, 2007), Lipunov \& Gorbovskoy (2007), 
Shafee et al. (2007) for more realistic situations.

Next, out of several plans, we would like to use our very approach to understand 
the physics of the symbiotic connection of disk and jet around black holes which, in 
the astrophysical context, is called by jet-disk symbiosis. In this context, one can
extend the qualitative picture of the jet-disk symbiosis developed earlier (Falcke \& Biermann 1995, 
Donea \& Biermann 1996) to a full scale threedimensional MHD in the Newtonian framework with 
the inclusion of our PNP. The structure of the disk is highly 
modified in presence of the jet. Thus, we plan to describe the 
disk including all the above mentioned effects 
in a realistic situation, and to develop a tool to test with observation, both 
of the disk and the jet (from radio to TEV photons).

\acknowledgements
The authors thank the referee for his/her very useful and constructive suggestions which helped
to improve the presentation of the paper. They also thank Paul J. Wiita for his comments
in preparing the final version of the manuscript.

{}

\clearpage
\begin{table*}[htbp]
\scriptsize
\caption{Values of $r_b$}
\begin{tabular}{ccccccccccccccccccccccccccccccccccccccc}
\hline
\hline
$ $ & $ $&K&$ $ & $ $&$ $&K&$ $ & $ $&$ $&K&$ $ & $ $&$ $&K&$ $ & $ $  \\
$a$ & $i$&$ $& Er & $ $ &$i$&$ $& Er & $ $ &$i$&$ $& Er & $ $ &$i$&$ $&Er \\
$ $ & $ $&PK&$ $ & $ $&$ $&PK&$ $ & $ $&$ $&PK&$ $ & $ $&$ $&PK&$ $ & $ $  \\
\hline
\hline
$ $ & $ $ & $1.0$ &$ $ & $ $ & $ $ &$1.0$  &$ $ & $ $ & $ $ &$1.0$ &$ $ & $ $ & $ $
&$1.0$   \\
$1.0$ & $0$&$ $&$0.0$&$ $ &$10$ & $ $&$0.0$ &$ $ &$20$ & $ $&$0.0$  &$ $ &$30$ & $
$&$0.0$  \\
$ $ & $ $ & $1.0$&$ $ & $ $ & $ $ &$1.0$&$ $ & $ $ & $ $ &$1.0$ &$ $ & $ $ & $ $ &$1.0$
\\
\hline
$ $ & $ $ & $1.0914$ &$ $ & $ $ & $ $ &$1.0942$  &$ $ & $ $ & $ $ &$1.1048$ &$ $ & $ $ & $ $
&$1.1369$   \\
$0.998$ & $0$&$ $&$5.0119$&$ $ &$10$ & $ $&$5.1088$ &$ $ &$20$ & $ $&$5.4671$  &$ $ &$30$ & $
$&$6.6233$  \\
$ $ & $ $ & $1.0367$&$ $ & $ $ & $ $ &$1.0383$&$ $ & $ $ & $ $ &$1.0444$ &$ $ & $ $ & $ $ &$1.0616$
\\
\hline
$ $ & $ $ & $1.7325$ &$ $ & $ $ & $ $ &$1.7459$  &$ $ & $ $ & $ $ &$1.7890$ &$ $ & $ $ & $ $
&$1.8706$   \\
$0.9$ & $0$&$ $&$4.6118$&$ $ &$10$ & $ $&$4.5535$ &$ $ &$20$ & $ $&$4.3823$  &$ $ &$30$ & $
$&$4.0468$  \\
$ $ & $ $ & $1.6526$&$ $ & $ $ & $ $ &$1.6664$&$ $ & $ $ & $ $ &$1.7106$ &$ $ & $ $ & $ $ &$1.7949$
\\
\hline
$ $ & $ $ & $2.3955$ &$ $ & $ $ & $ $ &$2.4106$  &$ $ & $ $ & $ $ &$2.457$ &$ $ & $ $ & $ $ &$2.5369$
\\
$0.7$ & $0$&$ $&$2.5924$&$ $ &$10$ & $ $&$2.5429$ &$ $ &$20$ & $ $&$2.4217$  &$ $ &$30$ & $
$&$2.1838$  \\
$ $ & $ $ & $2.3334$&$ $ & $ $ & $ $ &$2.3493$&$ $ & $ $ & $ $ &$2.3975$ &$ $ & $ $ & $ $ &$2.4815$
\\
\hline
$ $ & $ $ & $2.9142$ &$ $ & $ $ & $ $ &$2.9268$  &$ $ & $ $ & $ $ &$2.9648$ &$ $ & $ $ & $ $
&$3.0281$   \\
$0.5$ & $0$&$ $&$1.5064$&$ $ &$10$ & $ $&$1.4760$ &$ $ &$20$ & $ $&$1.3964$  &$ $ &$30$ & $
$&$1.2615$  \\
$ $ & $ $ & $2.8703$&$ $ & $ $ & $ $ &$2.8836$&$ $ & $ $ & $ $ &$2.9234$ &$ $ & $ $ & $ $ &$2.9899$
\\
\hline
$ $ & $ $ & $3.3733$ &$ $ & $ $ & $ $ &$3.3816$  &$ $ & $ $ & $ $ &$3.4064$ &$ $ & $ $ & $ $
&$3.4473$   \\
$0.3$ & $0$&$ $&$0.7826$&$ $ &$10$ & $ $&$0.7541$ &$ $ &$20$ & $ $&$0.7134$  &$ $ &$30$ & $
$&$0.6469$  \\
$ $ & $ $ & $3.3469$&$ $ & $ $ & $ $ &$3.3561$&$ $ & $ $ & $ $ &$3.3821$ &$ $ & $ $ & $ $ &$3.4250$
\\
\hline
$ $ & $ $ & $3.7974$ &$ $ & $ $ & $ $ &$3.8003$  &$ $ & $ $ & $ $ &$3.8091$ &$ $ & $ $ & $ $
&$3.8235$   \\
$0.1$ & $0$&$ $&$0.2317$&$ $ &$10$ & $ $&$0.2079$ &$ $ &$20$ & $ $&$0.1969$  &$ $ &$30$ & $
$&$0.1805$  \\
$ $ & $ $ & $3.7886$&$ $ & $ $ & $ $ &$3.7924$&$ $ & $ $ & $ $ &$3.8016$ &$ $ & $ $ & $ $ &$3.8166$
\\
\hline
$ $ & $ $ & $4$ &$ $ & $ $ & $ $ &$4$  &$ $ & $ $ & $ $ &$4$ &$ $ & $ $ & $ $ &$4$   \\
$0$ & $0$&$ $&$0$&$ $ &$10$ & $ $&$0$ &$ $ &$20$ & $ $&$0$  &$ $ &$30$ & $ $&$0$  \\
$ $ & $ $ & $4$&$ $ & $ $ & $ $ &$4$&$ $ & $ $ & $ $ &$4$ &$ $ & $ $ & $ $ &$4$    \\
\hline
$ $ & $ $ & $4.1976$ &$ $ & $ $ & $ $ &$4.1945$  &$ $ & $ $ & $ $ &$4.1853$ &$ $ & $ $ & $ $
&$4.1702$   \\
$-0.1$ & $0$&$ $&$0.2311$&$ $ &$10$ & $ $&$0.2265$ &$ $ &$20$ & $ $&$0.2150$  &$ $ &$30$ & $
$&$0.2038$  \\
$ $ & $ $ & $4.2073$&$ $ & $ $ & $ $ &$4.2040$&$ $ & $ $ & $ $ &$4.1943$ &$ $ & $ $ & $ $ &$4.1787$
\\
\hline
$ $ & $ $ & $4.5804$ &$ $ & $ $ & $ $ &$4.5706$  &$ $ & $ $ & $ $ &$4.5418$ &$ $ & $ $ & $ $
&$4.4949$   \\
$-0.3$ & $0$&$ $&$0.5960$&$ $ &$10$ & $ $&$0.5907$ &$ $ &$20$ & $ $&$0.5681$  &$ $ &$30$ & $
$&$0.5273$  \\
$ $ & $ $ & $4.6077$&$ $ & $ $ & $ $ &$4.5976$&$ $ & $ $ & $ $ &$4.5676$ &$ $ & $ $ & $ $ &$4.5186$
\\
\hline
$ $ & $ $ & $4.9495$ &$ $ & $ $ & $ $ &$4.9328$  &$ $ & $ $ & $ $ &$4.8831$ &$ $ & $ $ & $ $
&$4.8022$   \\
$-0.5$ & $0$&$ $&$0.9072$&$ $ &$10$ & $ $&$0.8981$ &$ $ &$20$ & $ $&$0.8683$  &$ $ &$30$ & $
$&$0.8142$  \\
$ $ & $ $ & $4.9944$&$ $ & $ $ & $ $ &$4.9771$&$ $ & $ $ & $ $ &$4.9255$ &$ $ & $ $ & $ $ &$4.8413$
\\
\hline
$ $ & $ $ & $5.3077$ &$ $ & $ $ & $ $ &$5.2836$  &$ $ & $ $ & $ $ &$5.2121$ &$ $ & $ $ & $ $
&$5.0953$   \\
$-0.7$ & $0$&$ $&$1.1775$&$ $ &$10$ & $ $&$1.1659$ &$ $ &$20$ & $ $&$1.1320$  &$ $ &$30$ & $
$&$1.0716$  \\
$ $ & $ $ & $5.3702$&$ $ & $ $ & $ $ &$5.3452$&$ $ & $ $ & $ $ &$5.2711$ &$ $ & $ $ & $ $ &$5.1499$
\\
\hline
$ $ & $ $ & $5.6568$ &$ $ & $ $ & $ $ &$5.6251$  &$ $ & $ $ & $ $ &$5.5308$ &$ $ & $ $ & $ $
&$5.3765$   \\
$-0.9$ & $0$&$ $&$1.4160$&$ $ &$10$ & $ $&$1.4044$ &$ $ &$20$ & $ $&$1.3687$  &$ $ &$30$ & $
$&$1.3038$  \\
$ $ & $ $ & $5.7369$&$ $ & $ $ & $ $ &$5.7041$&$ $ & $ $ & $ $ &$5.6065$ &$ $ & $ $ & $ $ &$5.4466$
\\
\hline
$ $ & $ $ & $5.8250$ &$ $ & $ $ & $ $ &$5.7895$  &$ $ & $ $ & $ $ &$5.6837$ &$ $ & $ $ & $ $
&$5.5104$   \\
$-0.998$ & $0$&$ $&$1.5227$&$ $ &$10$ & $ $&$1.5096$ &$ $ &$20$ & $ $&$1.4762$  &$ $ &$30$ & $
$&$1.4137$  \\
$ $ & $ $ & $5.9137$&$ $ & $ $ & $ $ &$5.8769$&$ $ & $ $ & $ $ &$5.7676$ &$ $ & $ $ & $ $ &$5.5883$
\\
\hline
$ $ & $ $ & $5.8284$ &$ $ & $ $ & $ $ &$5.7928$  &$ $ & $ $ & $ $ &$5.6868$ &$ $ & $ $ & $ $
&$5.5131$   \\
$-1.0$ & $0$&$ $&$1.4927$&$ $ &$10$ & $ $&$1.4794$ &$ $ &$20$ & $ $&$1.4419$  &$ $ &$30$ & $
$&$1.3731$  \\
$ $ & $ $ & $5.9154$&$ $ & $ $ & $ $ &$5.8785$&$ $ & $ $ & $ $ &$5.7688$ &$ $ & $ $ & $ $ &$5.5888$
\\
\hline
\hline
\end{tabular}
\tablecomments{Percentage error=Er. Inclination angle $i$ expressed in degrees,
K and PK denote Kerr and pseudo-Newtonian result respectively.}
\end{table*}

\clearpage
\begin{table*}[htbp]
\scriptsize
\caption{Values of $E_s$}
\begin{tabular}{ccccccccccccccccccccccccccccccccccccccc}
\hline
\hline
$ $ & $ $&K&$ $ & $ $&K&$ $ & $ $&K&$ $ & $ $&K&$ $ &$ $ \\
$a$ & $i$&$ $& Er & $i$&$ $& Er & $i$&$ $& Er & $i$&$ $&Er \\
$ $ & $ $&PK&$ $ & $ $&PK&$ $ & $ $&PK&$ $ & $ $&PK&$ $ & $ $ \\
\hline
\hline
$ $ & $ $ & $0.0$&$ $&$ $&$0.0$&$ $&$ $&$0.0$&$ $&$ $&$0.0$   \\
$1.0$ & $0$&$ $&$0.0$&$10$ & $ $&$0.0$ &$20$ & $ $&$0.0$  &$30$ & $ $&$0.0$  \\
$ $ & $ $ & $0.0$&$ $ & $ $ & $0.0$&$ $ & $ $ &$0.0$ &$ $ &$ $ & $0.0$    \\
\hline
$ $ & $ $ & $-0.3210$&$ $&$ $&$-0.3118$&$ $&$ $&$-0.2816$&$ $&$ $&$-0.2264$   \\
$0.998$ & $0$&$ $&$10.0623$&$10$ & $ $&$6.8954$ &$20$ & $ $&$2.1307$  &$30$ & $ $&$15.7243$  \\
$ $ & $ $ & $-0.3533$&$ $ & $ $ & $-0.3333$&$ $ & $ $ &$-0.2756$ &$ $ &$ $ & $-0.1908$    \\
\hline
$ $ & $ $ & $-0.1558$&$ $&$ $ &$-0.1530$  &$ $ & $ $ &$-0.1450$ &$ $ & $  $ &$-0.1327$   \\
$0.9$ & $0$&$ $&$12.1951$&$10$ & $ $&$8.8235$ &$20$ & $ $&$0.8966$  &$30$ & $ $&$15.9005$  \\
$ $ & $ $ & $-0.1748$&$ $ & $ $ &$-0.1665$&$ $ & $ $ &$-0.1437$ &$ $ & $ $ &$-0.1116$    \\
\hline
$ $ & $ $ & $-0.1036$ &$ $ & $ $ &$-0.1026$  &$ $ & $ $ &$-0.0995$ &$ $ & $ $ &$-0.0949$   \\
$0.7$ & $0$&$ $&$10.8108$&$10$ & $ $&$7.4074$&$20$ & $ $&$2.1106$  &$30$ & $ $&$15.9652$  \\
$ $ & $ $ & $-0.1148$&$ $ & $ $ &$-0.1102$&$ $ & $ $ &$-0.0974$ &$ $ & $ $ &$-0.0788$    \\
\hline
$ $ & $ $ & $-0.0821$ & $ $ & $ $ &$-0.0816$  & $ $ & $ $ &$-0.0802$ & $ $ & $ $ &$-0.0780$   \\
$0.5$ & $0$&$ $&$9.9757$&$10$ & $ $&$6.7402$&$20$ & $ $&$2.8678$  &$30$ & $ $&$17.5641$  \\
$ $ & $ $ & $-0.0904$&$ $ & $ $ &$-0.0871$&$ $ & $ $ &$-0.0779$ & $ $ & $ $ &$-0.0643$    \\
\hline
$ $ & $ $ & $-0.0694$ &$ $ & $ $ &$-0.0691$  &$ $ & $ $ &$-0.0685$ &$ $ & $ $ &$-0.0675$   \\
$0.3$ & $0$&$ $&$9.5101$&$10$ & $ $&$6.3676$ &$20$ & $ $&$3.2117$  &$30$ & $ $&$17.7778$  \\
$ $ & $ $ & $-0.0760$&$ $ &  $ $ &$-0.0735$&$ $ &  $ $ &$-0.0663$ &$ $ &  $ $ &$-0.0555$    \\
\hline
$ $ & $ $ & $-0.0606$ &$ $ & $ $ &$-0.0606$  &$ $ & $ $ &$-0.0604$ &$ $ & $ $ &$-0.0602$   \\
$0.1$ & $0$&$ $&$9.2409$&$10$ & $ $&$5.9406$ &$20$ & $ $&$3.4768$  &$30$ & $ $&$18.1063$  \\
$ $ & $ $ & $-0.0662$&$ $ & $ $ &$-0.0642$&$ $ & $ $ &$-0.0583$ &$ $ & $ $ &$-0.0493$    \\
\hline
$ $ & $ $ & $-0.0572$ &$ $ & $ $ &$-0.0572$  &$ $ & $ $ &$-0.0572$ & $ $ & $ $ &$-0.0572$   \\
$0$ & $0$&$ $&$9.0909$&$10$ & $ $&$5.9441$&$20$ & $ $&$3.4965$  &$30$ & $ $&$18.1818$  \\
$ $ & $ $ & $-0.0624$&$ $ & $ $ &$-0.0606$&$ $ & $ $ &$-0.0552$ &$ $ & $ $ &$-0.0468$    \\
\hline
$ $ & $ $ & $-0.0542$ &$  $ & $ $ &$-0.0542$  &$ $ & $ $ &$-0.0544$ &$ $ & $ $ &$-0.0546$   \\
$-0.1$ & $0$&$ $&$9.0406$&$10$ & $ $&$5.7196$ &$20$ & $ $&$3.8603$  &$30$ & $ $&$18.3150$  \\
$ $ & $ $ & $-0.0591$&$ $ & $ $ &$-0.0573$&$ $ & $ $ &$-0.0523$ &$ $ & $ $ &$-0.0446$    \\
\hline
$ $ & $ $ & $-0.0492$ &$ $ & $ $ &$-0.0493$  &$ $ & $ $ &$-0.0496$ &$ $ & $ $ &$-0.0502$   \\
$-0.3$ & $0$&$ $&$8.7398$&$10$ & $ $&$5.4767$ &$20$ & $ $&$3.8306$ &$30$ & $ $&$18.5259$  \\
$ $ & $ $ & $-0.0535$&$ $ & $ $ &$-0.0520$&$ $ & $ $ &$-0.0477$ &$ $ & $ $ &$-0.0409$    \\
\hline
$ $ & $ $ & $-0.0451$ &$ $ & $ $ &$-0.0453$  &$ $ & $ $ &$-0.0458$ &$ $ & $ $ &$-0.0465$   \\
$-0.5$ & $0$&$ $&$8.8647$&$10$ & $ $&$5.2980$ &$20$ & $ $&$4.1485$  &$30$ & $ $&$18.4946$  \\
$ $ & $ $ & $-0.0490$&$ $ & $ $ &$-0.0477$&$ $ & $ $ &$-0.0439$ &$ $ & $ $ &$-0.0379$    \\
\hline
$ $ & $ $ & $-0.0418$ &$ $ & $ $ &$-0.0420$  &$ $ & $ $ &$-0.0426$ &$ $ & $ $ &$-0.0435$   \\
$-0.7$ & $0$&$ $&$8.3732$&$10$ & $ $&$5.2381$ &$20$ & $ $&$4.2254$  &$30$ & $ $&$18.6207$  \\
$ $ & $ $ & $-0.0453$&$ $ & $ $ &$-0.0442$&$ $ & $ $ &$-0.0408$ &$ $ & $ $ &$-0.0354$    \\
\hline
$ $ & $ $ & $-0.0390$ &$ $ & $ $ &$-0.0392$  &$ $ & $ $ &$-0.0398$ &$ $ & $ $ &$-0.0409$   \\
$-0.9$ & $0$&$ $&$8.2051$&$10$ & $ $&$5.1020$ &$20$ & $ $&$4.2714$  &$30$ & $ $&$18.5819$  \\
$ $ & $ $ & $-0.0422$&$ $ & $ $ &$-0.0412$&$ $ & $ $ &$-0.0381$ &$ $ & $ $ &$-0.0333$    \\
\hline
$ $ & $ $ & $-0.0378$ &$ $ & $ $ &$-0.0380$  &$ $ & $ $ &$-0.0387$ &$ $ & $ $ &$-0.0398$   \\
$-0.998$ & $0$&$ $&$8.4656$&$10$ & $ $&$5.2632$ &$20$ & $ $&$3.8760$  &$30$ & $ $&$17.8392$  \\
$ $ & $ $ & $-0.0410$&$ $ & $ $ &$-0.0400$&$ $ & $ $ &$-0.0372$ &$ $ & $ $ &$-0.0327$    \\
\hline
\hline
$ $ & $ $ & $-0.0377$ &$ $ & $ $ &$-0.0380$  &$ $ & $ $ &$-0.0386$ &$ $ & $ $ &$-0.0398$   \\
$-1.0$ & $0$&$ $&$8.7533$&$10$ & $ $&$5.2632$ &$20$ & $ $&$3.8860$  &$30$ & $ $&$17.8392$  \\
$ $ & $ $ & $-0.0410$&$ $ & $ $ &$-0.0400$&$ $ & $ $ &$-0.0371$ &$ $ & $ $ &$-0.0327$    \\
\hline
\end{tabular}
\tablecomments{Percentage error=Er. Inclination angle $i$ expressed in degrees,
K and PK denote Kerr and pseudo-Newtonian result respectively.}
\end{table*}

\clearpage

\clearpage

\clearpage

\clearpage

\clearpage

\end{document}